\documentclass[aps,prb,twocolumn,showpacs,groupedaddress,graphicx]{revtex4}
\usepackage[dvips]{graphicx}
\usepackage{dcolumn}
\usepackage{amsmath}
\newcommand{\be}{\begin{equation}}
\newcommand{\ee}{\end{equation}}

\newcommand{\bea}{\begin{eqnarray}}
\newcommand{\eea}{\end{eqnarray}}
\newcommand{\bd}{\begin{displaymath}}
\newcommand{\ed}{\end{displaymath}}
\newcommand{\bi}{\begin{itemize}}
\newcommand{\ei}{\end{itemize}}
\newcommand{\bc}{\begin{center}}
\newcommand{\ec}{\end{center}}
\newcommand{\bfl}{\begin{flushleft}}
\newcommand{\efl}{\end{flushleft}}
\newcommand{\bfr}{\begin{flushright}}
\newcommand{\efr}{\end{flushright}}
\newcommand{\f}{\frac}

\def\bk{{\bf k}}  \def\bp{{\bf p}}

\def\6{\partial}  
\def\g{\gamma}

\def\o{\omega}  \def\D{\Delta}
  
  \def\O{\Omega}

\def\={\!\!\!&=&\!\!\!}
\def\+{\!\!\!&&\!\!\!+~}
\def\-{\!\!\!&&\!\!\!-~}

\begin{document}

\title{Specific heat behavior of high temperature superconductors in the pseudogap regime}

\author{Ionel \c{T}ifrea}

\affiliation{Department of Physics and Astronomy, University of
Iowa, Iowa City, Iowa 52242, USA}

\affiliation{Department of Theoretical Physics, "Babe\c s-Bolyai"
University, 3400 Cluj, Romania}

\author{C\u at\u alin  Pa\c scu Moca}

\affiliation{Department of Physics, University of Oradea, 3700
Oradea, Romania}

\date{\today}

\begin{abstract}
Experimental data obtained from thermodynamic measurements in
underdoped high temperature superconductors show unusual anomalies
in the temperature dependence of the electronic specific heat both
in the normal state and at the critical point associated to the
superconducting phase transition. The observed deviations from the
standard behavior are probably linked with the opening of a
pseudogap in the energy spectrum of the single-particle
excitations associated with the normal state. Based on a
phenomenological description of the pseudogap phase we perform
analytical and numerical calculations for the temperature
dependence of the specific heat for both the superconducting and
normal state. The reduced specific heat jump at the transition
point can be explained by a modified electronic single particle
contribution to the specific heat in the presence of the normal
state pseudogap. The hump observed in the normal state specific
heat can be explained by the electronic pair contribution
associated with strong fluctuations of the order parameter in the
critical region. The obtained theoretical results are discussed in
connection with experimental data for cuprates.
\end{abstract}

\maketitle

\section{Introduction}

One of the most controversial properties of high temperature
superconductor materials (HTSC) is the presence of a gap in their
normal state single particle excitation spectrum.\cite{pgaprev}
Usually addressed as the pseudogap, this relatively new feature
just added a new controversy on the long list of unusual
properties of the normal phase of HTSC. The pseudogap phase is
seen in the underdoped region of the HTSC phase diagram for
temperature values above the superconducting critical temperature
$T_c$ and below a characteristic temperature $T^*$. The presence
of the pseudogap phase was experimentally proved by direct
measurements of the single particle excitation spectra in angle
resolved photoemission spectroscopy (ARPES)\cite{arpes1,arpes2}
and tunnelling experiments\cite{tunelare}, but also in nuclear
magnetic resonance (NMR), specific heat, resistivity, infrared
conductivity, and Raman spectroscopy experiments.\cite{tallon}
However, in spite of a large amount of experimental data, there is
still no general consensus on the nature of the pseudogap phase,
especially regarding the doping dependence of the onset
temperature $T^*$ around the optimal doping point. As a function
of doping, in one possible scenario, $T^*$ merges with $T_c$ in
the overdoped region, whereas in a different scenario, $T^*$ falls
from large values in the underdoped region to a zero value at a
critical point, universally identify for all HTSC at
$p_{cr}=0.19$.\cite{tallon} From the theoretical point of view,
these two scenarios involve different approaches. In the first
one, the superconducting and pseudogap phases are strongly
connected, the pseudogap being associated with the formation of
precursor Cooper pairs ($T_c<T<T^*$), pairs which become coherent
and condense at $T_c$, leading to the superconducting
phase.\cite{crossover1,crossover2} A different theoretical
approach leading to the same conclusions with respect to the onset
temperature $T^*$ consider the role of the pair fluctuations above
the critical temperature $T_c$.\cite{pair1,pair2,pair3} In the
second scenario, the key role is played by the presence of an
antiferromagnetic region in the phase diagram at low doping
values, the pseudogap being a consequence of the direct
interaction between the electrons and fluctuations of the
antiferromagnetic order parameter.\cite{antifero1,antifero2}
Unfortunately, despite this large theoretical effort, an agreed
description of the normal state in HTSC (including the pseudogap
phase) is still lacking.

In a recent work, Moca and Janko,\cite{moca} performed a detailed
theoretical analysis of the electronic specific heat in HTSC.
Starting from a phenomenological description of the pseudogap, it
is argued that a correct description of the specific heat behavior
in the normal state can be obtained only with the inclusion of the
electron pair contribution. Such a scenario is sustained by the
presence of strongly enhanced fluctuations of the order parameter
in the critical region above the transition temperature in
quasi-two dimensional systems such as HTSC. In this way the
observed maximum\cite{loram2,loram} in the coefficient of the
electronic heat capacity, $\gamma(T)=C/T$, can be fully explained.
However, the anomalies related to the specific heat behavior are
observed also in the specific heat coefficient jump, $\Delta\gamma
(T_c)$, at the transition point.\cite{tallon} In the overdoped
region, where the presence of a pseudogap is still questionable,
the specific heat jump remains almost constant. As the doping
decreases, around the optimal doping point, where the pseudogap is
supposed to open, the specific heat jump starts to decrease. Once
the doping value is in the underdoped region, $\Delta\gamma (T_c)$
falls sharply, the larger the pseudogap is, the smaller the
specific heat jump becomes. The goal of this paper is to calculate
the temperature dependence of the electronic specific heat, both
for the superconducting and normal phase of HTSC, and to analyze
its jump at the critical point based on a modified BCS theory
which includes the pseudogap effects. A similar approach of the
electronic specific heat behavior was considered by Loram {\em et
al.}\cite{loram2} in order to explain the anomalous properties
induced by the pseudogap in YBa$_2$Cu$_3$O$_{6+x}$.

The paper is organized as follows: In Section II we discuss a
phenomenological theoretical approach based on the validity of
Gorkov's equations and we obtain the modified gap equation
starting from a normal state characterized by the presence of a
pseudogap, $E_g$. In Section III we discuss the specific heat
behavior based on the proposed model. Analytical results are
obtained for the specific heat jump at the critical point using
Pauli's theorem. Numerical results for the specific heat
coefficient, $\gamma(T)$, are presented based on an analytical
expression of the free energy as function of temperature, below
and above the critical temperature $T_c$. A comparison between
analytical and numerical results for the specific heat jump at the
transition point is also presented. Section IV gives our
conclusions.

\section{Theoretical model}

In the following we will consider a simple model based on the
Gorkov's equations formalism in which the normal state Green's
functions will include the presence of the pseudogap. A similar
analysis was used by different authors\cite{model1,model2,model3}
in order to study the effect of the pseudogap phase on different
properties of the superconducting state. Our theoretical approach
is based on the assumption that in the pseudogap phase the
self-energy corrections to the free electronic Green's function
are given by:

\begin{equation}\label{selfenergy}
{\Sigma}({\mathbf k}, i{\omega}_n)=-E^2_g({\mathbf k})
G_0(-{\mathbf k}, -i{\omega}_n)\; ,
\end{equation}
where $G_0(-{\mathbf k},-i{\omega}_n)$ represents the free
electron Green's function, $E_g({\mathbf k})$ the pseudogap, and
${\omega}_n=(2n+1)\pi T$ is the usual fermionic Matsubara
frequency. This phenomenological form of the self-energy was
already used to explain the form of the spectral function, $A(k_F,
{\omega})$, observed in ARPES experiments.\cite{oleg} A similar
behavior of the electronic self-energy was reported late in the
seventies by Schmid\cite{schmid} as a direct consequence of
electron-pair fluctuation interactions in the critical region
around the transition temperature. However, it is not our goal to
understand the origin of this phenomenological self-energy, but to
use it in order to extract different properties of the
superconducting state in cuprates. The normal state Green's
function can be obtained with the aid of Dyson's equation as:
\begin{equation}\label{greenf}
G({\mathbf k}, i{\omega}_n)=\frac{u^2_{\mathbf
k}}{i{\omega}_n-E_{\mathbf k}}+\frac{v^2_{\mathbf
k}}{i{\omega}_n+E_{\mathbf k}}\; ,
\end{equation}
where $E^2_{\mathbf k}=\xi^2_{\mathbf k}+E^2_g({\mathbf k})$,
$u^2_{\mathbf k}=(1+\xi_{\mathbf k}/E_{\mathbf k})/2$, and
$v^2_{\mathbf k}=(1-\xi_{\mathbf k}/E_{\mathbf k})/2$
($\xi_{\mathbf k}$ denotes the electron energy measured from the
Fermi level).

In terms of Green's function formalism the standard BCS theory is
recovered by the use of the Gorkov equations:
\begin{eqnarray}\label{gorkov}
&&G^{-1}_0({\mathbf k}, i{\omega}_n){\cal G}({\mathbf k},
i{\omega}_n)+\Delta({\mathbf k}){\cal F}^\dagger({\mathbf k},
i{\omega}_n)=1\nonumber\\ &&\Delta^*({\mathbf k}){\cal G}({\mathbf
k}, i{\omega}_n)-G^{-1}_0(-{\mathbf k}, -i{\omega}_n){\cal
F}^\dagger({\mathbf k}, i{\omega}_n)=0\;.\;\;\;
\end{eqnarray}
${\cal G}({\mathbf k}, i{\omega}_n)$ and ${\cal F}({\mathbf k},
i{\omega}_n)$ represent the normal and anomalous Green's functions
in the superconducting state. The superconducting order parameter,
$\Delta({\mathbf k})$, is defined in the usual way in terms of the
anomalous Green's function, ${\cal F}({\mathbf k}, i{\omega}_n)$,
as:
\begin{equation}\label{gap}
\Delta({\mathbf k})=-T\sum_n\int\frac{d{\mathbf
p}}{(2\pi)^2}V({\mathbf k}, {\mathbf p}) {\cal F}^\dagger({\mathbf
p}, i{\omega}_n)\;.
\end{equation}
The interaction term, $V({\mathbf k}, {\mathbf p})$, supposed to
be attractive, is responsible for the formation of the Cooper
pairs. The anomalous superconducting state Green's function can be
easily obtained from Gorkov's equations and using Eq. (\ref{gap})
the standard BCS gap equation is recovered.

Our theoretical model assumes the validity of Gorkov's equations
for the case of HTSC, where the free electron Green's function,
$G_0({\mathbf k}, i{\omega}_n)$, is replaced by the more general
Green's function given by Eq. (\ref{greenf}). In this way the
effects of the pseudogap on the superconducting gap equation are
considered. A simple calculation leads to the following general
gap equation:
\begin{eqnarray}\label{delta}
&&\Delta({\mathbf k})=-T\sum_n\int\frac{d{\mathbf
p}}{(2\pi)^2}V({\mathbf k}, {\mathbf p})\nonumber\\
&&\times\frac{\Delta({\mathbf
p})\left[(i{\omega}_n)^2-\xi_{\mathbf p}^2\right]}
{|\Delta({\mathbf p})|^2\left[(i{\omega}_n)^2-\xi_{\mathbf
p}^2\right]- \left[(i{\omega}_n)^2-\xi_{\mathbf
p}^2-E_g^2({\mathbf p})\right]^2}\;.\nonumber\\
\end{eqnarray}
The sum over the Matsubara frequencies can be performed
analytically and the gap equation becomes:
\begin{eqnarray}\label{gapfinal}
&&\frac{1}{V_d}=\int\frac{d{\mathbf
p}}{(2\pi)^2}\frac{\psi^2({\mathbf
p})}{2\sqrt{\Delta^4(T)+4\Delta^2(T)E^2_g}}
\nonumber\\&&\times\left[\frac{A^2(T)}{\sqrt{\xi^2_{\mathbf
p}+A^2(T)\psi^2({\mathbf p})}} \tanh{\frac{\sqrt{\xi^2_{\mathbf
p}+A^2(T)\psi^2({\mathbf p})}}{2T}}\right.\nonumber\\
&&-\left.\frac{B^2(T)}{\sqrt{\xi^2_{\mathbf
p}+B^2(T)\psi^2({\mathbf p})}} \tanh{\frac{\sqrt{\xi^2_{\mathbf
p}+B^2(T)\psi^2({\mathbf p})}}{2T}}\right]\;,\nonumber\\
\end{eqnarray}
where
\begin{eqnarray}\label{notations}
A^2(T)=E_g^2+\frac{1}{2}\left[\Delta^2(T)+\sqrt{\Delta^4(T)+4\Delta^2(T)E^2_g}\right]\;,\nonumber\\
B^2(T)=E_g^2+\frac{1}{2}\left[\Delta^2(T)-\sqrt{\Delta^4(T)+4\Delta^2(T)E^2_g}\right]\;.
\end{eqnarray}
$\psi({\mathbf p})$ in the gap equation (\ref{gapfinal}) is a
factor associated with the general symmetry properties of the
superconducting gap, pseudogap, and interaction potential.
Experimental data from ARPES and tunnelling experiments show that
both the superconducting gap and pseudogap have the same symmetry,
which in the case of HTSC is considered to be of $d$-wave
type,\cite{arpes1,arpes2} with $\psi({\mathbf
p})=\cos(2\theta_{\mathbf p})$ ($\theta_{\mathbf
p}=\arctan(p_y/p_x)$). Implicitly, the symmetry of the interaction
term is assumed to be of the same type, i.e, $V({\mathbf p},
{\mathbf k})=V_d\psi({\mathbf k})\psi({\mathbf p})$. The pure BCS
case is simply recovered in the $E_g{\rightarrow} 0$ limit. Note
at this point the differences between our general gap equation
(\ref{gapfinal}) and the one used by Loram {\em et
al.}\cite{loram2} in their approach. First of all, Eq.
(\ref{gapfinal}) includes contributions beyond the ones considered
in Ref. \onlinecite{loram2}, which can be seen as a particular
limit of Eq. (\ref{gapfinal}) for small values of the pseudogap,
$E_g$.\cite{jose} Secondly, we considered a $d$-wave symmetry of
the order parameter, which is known to be more appropriate for the
case of HTSC.\cite{arpes1,arpes2}

\begin{figure}[tbp]
\centering \scalebox{0.4}[0.45]{\includegraphics*{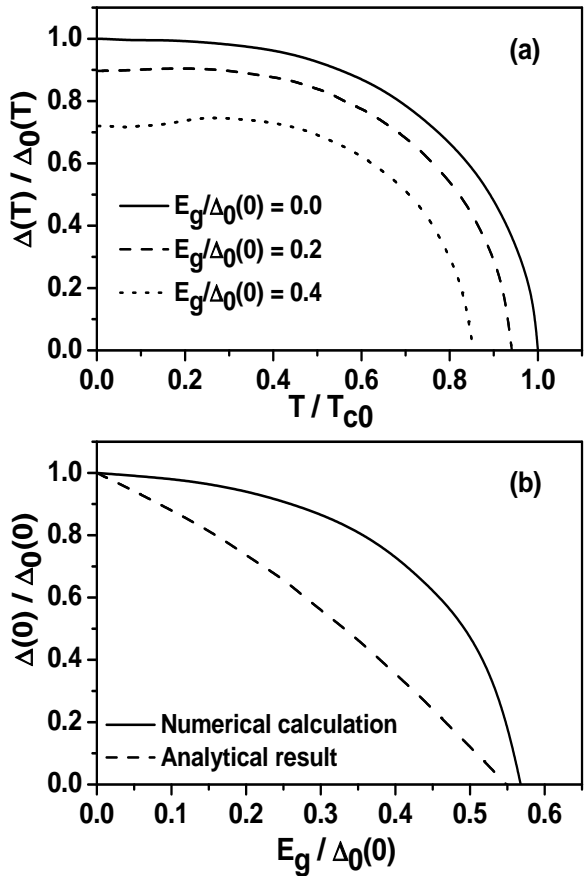}}
\caption{ (a) Numerical results for the temperature dependence of
the superconducting gap function for different values of the ratio
$E_g/\D_0(0)$ ($\D_0(0)$ represents the superconducting gap at
$T=0$ in the absence of the pseudogap). (b) The zero temperature
normalized superconducting gap, $\Delta (0)/\D_0(0)$, as function
of the ratio $E_g/\D_0(0)$ obtained from both the numerical (full
line) and analytical (dashed line) calculations.}
\label{Gap}
\end{figure}

Unfortunately, the exact analytical solution for the gap equation
(\ref{gapfinal}) at any temperature in the superconducting state
($0<T<T_c$) is difficult to obtain. Instead we performed a
numerical calculation for the gap temperature dependence
considering different values of the normal state pseudogap. As a
general observation we note that the superconducting state is
suppressed by the presence of the normal state pseudogap. The
value of the superconducting gap at $T=0$ K, $\D(0)$, transition
temperature, $T_c$, and superconducting gap, $\D(T)$, decrease as
the pseudogap increases. As an example of the numerical solution,
in Fig. \ref{Gap}(a) we present the temperature dependance of the
superconducting order parameter, $\D(T)$, as function of
temperature for different values of the pseudogap, $E_g$.

However, some simplifications in Eq. (\ref{gapfinal}) can be made
in order to approximate some of the superconducting state
properties, such as the value of the critical temperature, $T_c$,
and the superconducting gap at $T=0$, $\Delta(0)$. First of all
the integration over the momenta is replaced by an energy
integration using the corresponding density of states (DOS) in the
normal state:\cite{imp,model3}
\begin{equation}\label{DOS}
N(\xi)=\left\{\begin{array}{ll}\frac{1}{2}N_0\frac{|\xi|}{E_g}, &
|\xi|<E_g\\ \frac{1}{2}N_0, & |\xi|>E_g
\end{array}\right.\;,
\end{equation}
where $N_0$ denotes the DOS of a two dimensional (2D) free
electron gas. This form of the density of states resembles the one
seen experimentally in HTSC and is different from the constant
density of states used in Ref. \onlinecite{loram2}. We assume also
the following energy scale, $E_g<\Delta(0)<W/2$, with $W$ being
the bandwidth. This choice of the energy scale is valid in the
doping region around the optimal doping point.\cite{tallon} Based
on this assumption one finds:\cite{model3}
\begin{eqnarray}\label{tcd0}
T_c=T_{c0}\left[1-\frac{1}{4}\frac{E_g}{T_{c0}}-\frac{21\zeta(3)}{16\pi^2}
\left(\frac{E_g}{T_{c0}}\right)^2\right]\;,\nonumber\\
\Delta(0)=\Delta_0(0)\left[1-\frac{E_g}{\Delta_0(0)}-
\frac{3}{2}\left(\frac{E_g}{\Delta_0(0)}\right)^2\right]\;\label{Delta},
\end{eqnarray}
with $T_{c0}$ and $\Delta_0(0)$ being the critical temperature and
zero temperature order parameter obtained in standard $d$-wave BCS
calculations.\cite{dBCS} As we can see, the effect of the
pseudogap is to decrease both the critical temperature and the
superconducting gap at $T=0$ (See Ref. \onlinecite{model3}), in
agreement with the numerical results. Fig. \ref{Gap}(b) compares
the analytical result of Eq. (\ref{Delta}) for the pseudogap
dependence of the $T=0$ K order parameter with the exact numerical
result obtained by solving the gap equation (\ref{gapfinal}). The
slopes of the curves are slightly different, but such a behavior
is expected as for the analytical results a number of
simplifications were made.

\section{Specific heat}

One of the most systematic experimental studies of the
thermodynamic properties of the HTSC are the specific heat
measurements \cite{loram2,loram}. Anomalies in the specific heat
behavior are reported especially in the normal state, the
coefficient of the electronic specific heat capacity, $\gamma(T)$,
in underdoped samples being no longer constant as function of
temperature as we expect from the standard Fermi liquid theory.
The broad maximum observed in $\gamma(T)$ at a specific
temperature $T_m\sim T^*$ is associated with the onset of the
pseudogap. Moreover, the specific heat jump at the transition
point is a function of doping, a lower value than the standard one
predicted by the BCS theory being observed in the underdoped
region.

In this section we calculate the temperature dependence of the
specific heat below and above the critical temperature, extract
the specific heat jump at the critical point and compare our
analytical and numerical results with the available experimental
data.

\subsection{Pauli's theorem}

According to the Pauli's theorem the difference between the
superconducting and normal state thermodynamic potential at the
transition point can be calculated based on the following
formula:\cite{fetter}
\be\label{thermpot}
\f{\O_s-\O_n}{v}=\int_0^\D\;d\D'\left(\D'\right)^2\f{d(1/V_d)}{d\D'}\;,
\ee
where $\O_s$ and $\O_n$ are the thermodynamic potentials in the
superconducting and normal state respectively, and $v$ is the
sample volume. In the critical region the direct dependence
between the inverse of the interacting potential, $1/V_d$, and the
superconducting gap, $\D(T)$, can be extracted from Eq.
(\ref{delta}) considering that close to the transition point the
value of the order parameter is small. Accordingly, one finds that
\bea\label{deltacrit}
&&\f{1}{V_d}=-T\sum_n\int\f{d\bp}{(2\pi)^2}
\f{\psi^2(\bp)\left[\left(i\o_n\right)^2-\xi^2_\bp\right]}
{\left[\left(i\o_n\right)^2-\xi^2_\bp-E^2_g\psi^2(\bp)\right]^2}\nonumber\\
&&-\D^2(T)\;T\sum_n\int\f{d\bp}{(2\pi)^2}
\f{\psi^4(\bp)\left[\left(i\o_n\right)^2-\xi^2_\bp\right]^2}
{\left[\left(i\o_n\right)^2-\xi^2_\bp-E^2_g\psi^2(\bp)\right]^4}\;,\nonumber\\
\eea
a result which together with Eq. (\ref{thermpot}) leads to
\bea\label{pottherm}
&&\f{\O_s-\O_n}{v}=\nonumber\\
&&-\f{\D^4(T)}{2}\;T\sum_n\int\f{d\bp}{(2\pi)^2}
\f{\psi^4(\bp)\left[\left(i\o_n\right)^2-\xi^2_\bp\right]^2}
{\left[\left(i\o_n\right)^2-\xi^2_\bp-E^2_g\psi^2(\bp)\right]^4}\;.\nonumber\\
\eea
The electronic contribution to the specific heat can be calculated
based on the thermodynamic potential as $C=-T\6^2\O/\6^2T$. One
can see that the knowledge of the specific heat jump at the
critical point requires the knowledge of the temperature
dependence of the superconducting gap in the critical region,
below the phase transition point. A laborious, but straightforward
calculation based on Eq. (\ref{deltacrit}) gives
\bea\label{deltaT}
\D^2(T)&=&\f{32\pi^2T_c^2}{21\zeta(3)}\left(1-\f{T}{T_c}\right)\nonumber\\
&&\times\left[1-3.14\f{E_g}{\D_0(0)}+7.6\left(\f{E_g}{\D_0(0)}\right)^2\right]\;,\nonumber\\
\eea
a result which inserted in Eq. (\ref{pottherm}) gives for the
specific heat jump at the transition point the following value
\bea\label{analiticalcv}
&&\f{C_s-C_n}{v}=\nonumber\\&&\f{8\pi^2T_cN_0}{21\zeta(3)}
\f{\left[1-0.53\f{E_g}{\D_0(0)}-1.46\left(\f{E_g}{\D_0(0)}\right)^2\right]^2}
{1+2.62\f{E_g}{\D_0(0)}-2.21\left(\f{E_g}{\D_0(0)}\right)^2}\;.
\eea
It is clear from Eq. (\ref{analiticalcv}) that the presence of the
pseudogap in the single particle excitation spectrum in the normal
state is responsible for the observed suppression of the specific
heat jump at the critical point. Such an effect seems to be
universal, as it is observed in the underdoped region of most of
the HTSC materials phase diagram.\cite{tallon,loram2,loram}

\subsection{Numerical results}

The electronic single particle contribution to the specific heat
has a simple linear $T$-dependence above $T_c$ and an exponential
temperature dependence below the critical temperature. A detailed
numerical analysis of the temperature dependence of the specific
heat coefficient in underdoped HTSC was done in
Ref.\onlinecite{moca}. The main conclusion of this study was that
the electronic single particle contribution to the specific heat
is not enough to correctly describe the experimental data in the
normal state, the inclusion of the electronic pair contribution
being required in order to understand these data. The electronic
pair contribution, associated to strong fluctuations of the order
parameter in the critical region above the phase transition point,
induces a hump in the normal state specific heat at a temperature
of the order of $T^*$, and has to be considered to understand the
general behavior of the specific heat in the normal state of
HTSC.\cite{moca} However, no numerical results were presented for
the specific heat jump at the transition point.

In the following we will present numerical result for the
temperature dependence of the specific heat coefficient, $\g(T)$,
in both the superconducting and normal state. Our analysis will
include both electronic single particle and electronic pair
contributions. From the numerical point of view, a more reliable
way to extract the specific heat temperature dependence is to
compute $C$ as the first derivative of the electronic free energy,
$C=dE/dT$. In general the electronic free energy, $E$, can be
calculated based on the following relation:\cite{fetter}
\begin{equation}
E=T\sum\limits_{\mathbf{k},n}(i\omega _n+\xi _{\mathbf{k}}){\cal
G}(\mathbf{k} ,i\omega _n)\;.  \label{E}
\end{equation}
Below the transition temperature the normal Green's function
${\cal G}({\mathbf k } ,i\omega _n)$ can be obtained from Gorkov
equations, while for temperatures above the critical temperature
the normal state Green's function, $G(\mathbf{k} ,i\omega_n)$,
given by Eq (\ref{greenf}) should be used instead. Two steps are
performed in order to extract the temperature dependence of the
specific heat coefficient. First, we will sum over the Matsubara
frequencies and, secondly, we will numerically integrate over the
momentum space, $\mathbf k$. The general result can be written as:
\begin{equation}
E=\sum\limits_{\mathbf{k}}S_{\mathbf{k}}\;,
\end{equation}
where $S_{\mathbf{k}}$ has different values in the superconducting
and normal states. For the superconducting state, at temperature
smaller the critical temperature, $T_c$, after the summation over
the Matsubara frequencies is performed, $S_\bk$ can be calculated
from the following relation
\begin{eqnarray}
&&S_{\mathbf{k}}=\frac 1{\psi^2(\bk)\sqrt{\Delta(T)^4+4\Delta^2(T)E_g^2}}\nonumber\\
&&\times\left[ \frac{2\xi_{\mathbf{k}}\psi^2(\bk)\left(
A^2(T)-E_g^2\right)
\sqrt{\xi _{\mathbf{k}}^2+A(T)^2\psi^2(\bk)}}
{\sqrt{\xi_{\mathbf{k}}^2+A(T)^2\psi^2(\bk)}}\right. \nonumber \\
&&-\frac{2\xi _{\mathbf{k}}^2+A(T)^2\psi^2(\bk)}
{\sqrt{\xi _{\mathbf{k%
}}^2+A(T)^2\psi^2(\bk)}} \tanh{\f{\sqrt{\xi _{\mathbf{k}}^2+A(T)^2\psi^2(\bk)}}{2T}}
\nonumber \\
&&+\frac{2\xi_{\mathbf{k}}\psi^2(\bk)\left( B^2(T)-E_g^2\right)
\sqrt{\xi _{\mathbf{k}}^2+B(T)^2\psi^2(\bk)}}
{\sqrt{\xi_{\mathbf{k}}^2+B(T)^2\psi^2(\bk)}} \nonumber \\
&&\left.-\frac{2\xi _{\mathbf{k}}^2+B(T)^2\psi^2(\bk)}
{\sqrt{\xi _{\mathbf{k%
}}^2+B(T)^2\psi^2(\bk)}} \tanh{\f{\sqrt{\xi
_{\mathbf{k}}^2+B(T)^2\psi^2(\bk)}}{2T}}\right]\;,\nonumber\\
\label{Sk1}
\end{eqnarray}
whereas in for the normal state, at temperatures larger than the
critical temperature, $T_c$, one has:
\begin{eqnarray}
S_{\mathbf{k}} &=&\frac 1{\sqrt{\xi _{\mathbf{k}}^2+E_g^2\psi^2(\bk)}%
}\left[ 2\xi _{\mathbf{k}}\sqrt{\xi _{\mathbf{k}}^2+E_g^2\psi^2(\bk)}%
\right.   \nonumber \\
&&-\left( 2\xi _{\mathbf{k}}^2+E_g^2\psi^2(\bk)\right) \tanh{\f{\sqrt{%
\xi _{\mathbf{k}}^2+E_g^2\psi^2(\bk)}}{2T}}\;.\;   \label{Sk2}
\end{eqnarray}

\begin{figure}[tbp]
\centering
\scalebox{0.4}[0.45]{\includegraphics*{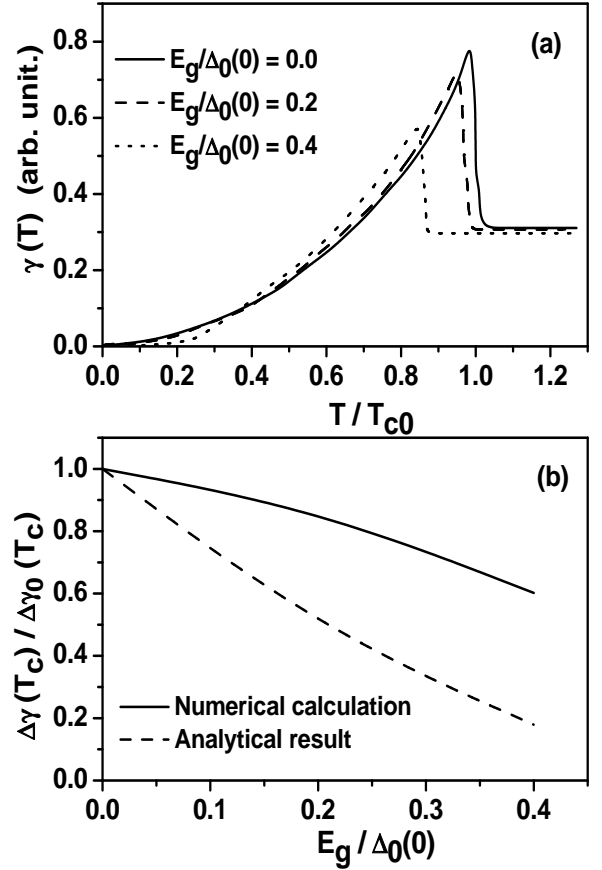}}
\caption{(a) Temperature dependence of the electronic single
particle contribution to the specific heat coefficient, $\g(T)$,
for different values of the ratio $E_g/\D_0(0)$. (b) The
normalized specific heat coefficient jump at the critical point,
$\D\g(T_c)/\D\g_0(T_c)$, ($\D\g_0(T_c)$ represents the value of
the specific heat coefficient jump at the transition point in the
absence of the pseudogap) as function of the ratio $E_g/\D_0(0)$
obtained from both the numerical (full line) and analytical
(dashed line) calculations.}
\label{SpecificHeatCoefficient}
\end{figure}

The results obtained for the temperature dependence of the total
energy, $E$, obtained by integrating  Eqs. (\ref{Sk1}) and
(\ref{Sk2}), match perfectly at the critical point but the slopes
of the curves are different, indicating a discontinuity of the
specific heat at the transition point. The specific heat
coefficient can be calculated in a straightforward manner by
taking the derivative as a function of temperature. The results
are presented in Fig. \ref{SpecificHeatCoefficient}(a). Two
important features are observed: first, as the analytical results
predicted, the specific heat coefficient jump at the critical
point decreases as the value of the pseudogap increases. Secondly,
one can see that in the normal state, above the transition
critical temperature, the pseudogap presence has no effect on the
specific heat coefficient, a constant value being obtained as the
electronic single particle contribution is considered. Therefore,
to explain the broad hump observed experimentally in the normal
state specific heat, different mechanisms should be considered.
Fig \ref{SpecificHeatCoefficient}(b) presents the normalized
specific heat coefficient jump at the transition point as function
of the ratio $E_g/\D_0(0)$, as extracted from numerical and
analytical calculations. As expected, there are slight differences
between the slopes of the numerical and analytical curves, a
feature which is due to successive approximations used in the
analytical calculation.

\begin{figure}[tbp]
\centering \scalebox{0.4}[0.45]{\includegraphics*{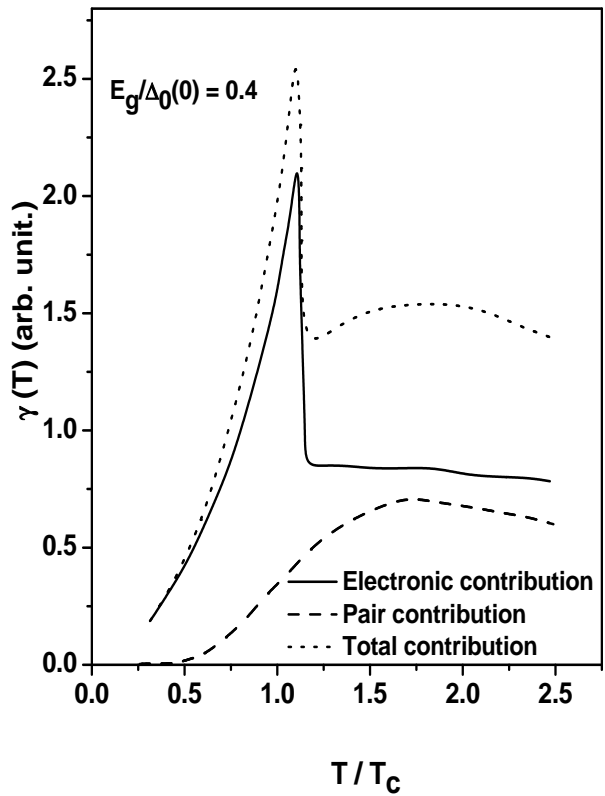}}
\caption{Different contributions to the specific heat coefficient
as function of temperature in the presence of the pseudogap
($E_g/\D_0(0)=0.4$). The total electronic contribution (dotted
line) to the specific heat coefficient is obtained as a sum of the
electronic single particle contribution (full line) and electronic
pair contribution (dashed line). A similar qualitative behavior
was reported for HTSC in Refs. \onlinecite{loram2,loram}.}
\label{SpecificHeat}
\end{figure}

As we already mentioned different contributions to the specific
heat coefficient, $\g(T)$, have to be considered for a better
understanding of the HTSC specific heat data at high temperatures
corresponding to the normal state.\cite{loram2,loram} One possible
mechanism was considered by Moca and Janko,\cite{moca} and is
related to the strong nature of the order parameter fluctuations
in the critical region, leading to an important electronic pair
contribution to the specific heat coefficient in the normal state.
This contribution has been evaluated previously \cite{moca} and we
quote here the result for completeness, $C^{pair}\sim (T^*/T)^2
\exp (-2T^*/T )$, where $T^*=2\pi E_g^2/g$ and $g$ describes the
quasiparticle interaction. Fig (\ref{SpecificHeat}) presents the
temperature dependance of the specific heat coefficient in the
presence of a normal state pseudogap ($E_g/\D(0)$). Electronic
single particle and electronic pair contributions are summed to
obtained the total electronic specific heat coefficient. The
electronic single particle contribution is important for the
specific heat coefficient jump at the transition point, whereas
the electronic pair contribution can explain the broad maximum
observed in the specific heat coefficient at high temperatures in
the normal state. Note that no specific heat coefficient jump at
the transition point is associated to the electronic pair
contribution. A similar qualitative behavior of the total
electronic specific heat coefficient was reported for
HTSC.\cite{tallon,loram2,loram}

\section{Conclusions}

The nature of the pseudogap is still an open question and its
explanation is beyond the scope of this paper. Using a simple
phenomenological model we have investigated the temperature
dependence of the specific heat coefficient both below and above
the critical point for HTSC characterized by the presence of a
pseudogap in the excitation spectrum of the normal phase. The
single particle contribution is the most important contribution to
the electronic specific heat, but this is not sufficient to
explain the hump that develops at high temperatures in the normal
state.

An important result that emerges from our calculations is the
behavior of the specific heat jump at the critical point.
Experimentally, in the underdoped region, where the pseudogap
energy is large, the jump is very small \cite{tallon} in contrast
with the overdoped region where the pseudogap is practically
absent and a BCS-like behavior emerges with a large jump at the
critical point. We have calculated the electronic single particle
contribution to the specific heat jump at the critical point in
the presence of the pseudogap and a similar dependence to the one
experimentally observed was obtained. Increasing the pseudogap the
specific heat jump at $T_c$ starts to decrease. This feature can
be understand in terms involving the suppression of the
superconducting state in the presence of the pseudogap, as both
the superconducting gap and critical temperature are smaller due
to a loss of states at the Fermi level. The specific heat anomaly
presented in HTSC at high temperatures ($T>T_c$) can be
successfully understood in terms of two particle (electronic pair)
contributions to the specific heat.\cite{moca} Note that the pair
contribution to the specific heat does not affect the specific
heat jump at the transition point.

Our analysis is based both on analytical and numerical
calculations. The restrictions imposed by the approximations used
in the analytical calculation lead to differences between the
analytical and numerical results. Basically, the analytical
results are reasonable around the optimal doping point and in the
overdoped region, where the value of the observed pseudogap is
small. With respect to the phase diagram one should mention that
the model we used is valid only in the second scenario discussed
in the introduction, as the pseudogap presence leads to a
reduction of the normal-superconducting phase transition
temperature, meaning that the pseudogap should completely
disappear around the optimal doping point. However, it is
generally accepted that in the overdoped region the standard Fermi
liquid and BCS theories are still valid, any inclusions of a
pseudogap in the description of the system properties being
inappropriate.


\end{document}